\documentclass[aps,prb,twocolumn,showpacs,letterpaper,superscriptaddress,longbibliography]{revtex4-1}

\usepackage[colorlinks=true, citecolor=blue, linkcolor=blue, urlcolor=blue]{hyperref}
\usepackage{graphicx,dcolumn,longtable,epsfig}
\usepackage[usenames]{color}
\usepackage{amssymb}
\usepackage{amsmath}
\usepackage{bm}
\usepackage{footnote}
\usepackage{float}
\usepackage{subfigure}
\usepackage{color}

\usepackage[english]{babel}

\usepackage{ulem}
\usepackage[T1]{fontenc}

\newcommand{\be}{\begin{equation}}
\newcommand{\ee}{\end{equation}}

\def\bea{\begin{eqnarray}}
\def\eea{\end{eqnarray}}

\begin{document}
\title{A comprehensive study of bond bipolaron superconductivity in triangular lattice}

\author{Chao Zhang}
\email{chaozhang@ahnu.edu.cn}
\affiliation{Department of Physics, Anhui Normal University, Wuhu, Anhui 241000, China}

\begin{abstract}

We employ the diagrammatic Monte Carlo method with a lattice path-integral formulation for both electron and phonon degrees of freedom to investigate the formation and properties of bond polarons and bipolarons on a two-dimensional triangular lattice. In the adiabatic regime ($\omega/t < 1.0$), single polarons remain light with a small effective mass, while bipolarons remain compact and lightweight, resulting in a high superfluid transition temperature $T_c$. We systematically study the dependence of $T_c$ on electron-phonon coupling strength and on-site interaction in the bond bipolaron model. Our results show that a moderate on-site repulsion enhances $T_c$ by stabilizing compact yet lightweight bipolarons, leading to high-$T_c$ superconductivity in the dilute limit. Notably, the triangular lattice sustains relatively high $T_c$ across a wide range of phonon frequencies, outperforming square lattice geometry. This enhancement arises from the higher coordination number and the bond-centered nature of the electron-phonon coupling. These findings suggest that triangular lattice geometry offers a promising platform for realizing high-$T_c$ bipolaronic superconductivity.

\end{abstract}

\maketitle

\section{Introduction}
\label{sec:sec1}

Polarons—quasiparticles formed when electrons dressed by lattice distortions—play a central role in determining the electronic properties of materials with strong electron-phonon interactions. Polaron effective mass and spatial extent critically affect transport and pairing~\cite{Millis1996,Bonca2000,Alexandrov:1999fy,Kornilovitch1998}, and may also shape collective behavior in systems dominated by electron-phonon interactions. In particular, the formation of bipolarons, where two polarons bind to form a composite boson, can lead to bipolaronic superconductivity when these pairs condense into a superfluid state~\cite{Alexandrov1996,
Devreese2020}. Achieving high-$T_c$ in such a scenario requires bipolarons that are both light and compact, a condition that depends sensitively on the underlying polaron properties~\cite{PhysRevX.13.011010}.

Among the proposed mechanisms for high-temperature superconductivity, bipolaronic pairing has attracted renewed attention~\cite{Alexandrov1996,Devreese2020}. A central challenge in this framework is achieving both strong pairing and sufficient mobility, as tightly bound bipolarons tend to be heavy and localized. This trade-off is particularly pronounced in the conventional Holstein model, where the electron-phonon coupling is purely on-site. Extensive numerical studies~\cite{PhysRevB.69.245111,PhysRevLett.84.3153,Chakraverty} have shown that Holstein bipolarons acquire large effective masses in the strong-coupling regime, severely suppressing the superconducting transition temperature $T_c$. Furthermore, at finite carrier density, the Holstein model tends to favor charge-density-wave (CDW) order over superconductivity~\cite{PhysRevLett.120.187003,PhysRevB.100.245105}, further limiting its ability to support coherent bipolaronic condensates. These limitations have motivated the exploration of alternative coupling mechanism that can support lighter, more mobile bipolarons.

An alternative strategy is to consider bond Su–Schrieffer–Heeger (SSH) electron-phonon interactions, where the coupling modulates the electronic hopping amplitudes between neighboring sites. Unlike the Holstein model, where the coupling is on-site and leads to heavy, localized bipolarons at strong coupling~\cite{PhysRevB.69.245111,PhysRevLett.84.3153,Chakraverty}, SSH-type interactions allow polarons to remain mobile even in the strong-coupling regime~\cite{PhysRevB.104.035143,PhysRevB.104.L140307}. This facilitates the formation of compact and lightweight bipolarons, which are favorable for achieving higher superfluid transition temperatures $T_c$~\cite{,PhysRevB.104.L201109,PhysRevB.108.L220502,PhysRevB.111.184513,PhysRevX.13.011010,PhysRevLett.121.247001,PhysRevB.109.L220502}.

Lattice geometry further plays a pivotal role in these phenomena in the bond SSH model. Triangular lattice, with their higher coordination number, offer increased hopping pathways for charge carriers. This enhanced connectivity can reduce the polaron effective mass and support the formation of compact, lightweight bipolarons in the bond SSH model. Despite these advantages, the properties of bond polarons and bipolarons on triangular lattice—particularly in the adiabatic and non-perturbative regimes—remain largely unexplored.

In this work, we employ diagrammatic Monte Carlo (DiagMC) simulations based on a lattice path-integral formulation for both electron and phonon sector, to investigate bond polarons and bipolarons on a two-dimensional triangular lattice~\cite{PhysRevB.105.L020501}. This unbiased numerical method enables us to systematically explore polaron and bipolaron formation across a wide parameter space without uncontrolled approximations. Our results show that in the adiabatic regime ($\omega/t < 1.0$), single polarons exhibit a small effective mass, while bipolarons remain compact and lightweight, leading to high superfluid transition temperatures $T_c/\omega$. Moreover, we find that the triangular lattice sustains relatively high $T_c$ values over a broader range of coupling strengths compared to the square lattice, owing to its enhanced coordination number and the bond-centered nature of the electron-phonon interaction. These findings suggest that triangular lattice geometry offers a promising route toward realizing high-$T_c$ bipolaronic superconductivity.

The remainder of this paper is organized as follows. In Sec.\ref{sec:sec2} and Sec.\ref{sec:sec3}, we introduce the bond SSH model on the triangular lattice and outline the DiagMC method used in this study. In Sec.\ref{sec:sec4}, we present results for single-polaron properties, including the ground-state energy, effective mass, quasiparticle residue, and average phonon number, and analyze how these quantities are influenced by the lattice geometry. In Sec.\ref{sec:sec5}, we turn to the bipolaron sector, examining the effective mass, mean-squared radius, and the resulting superfluid transition temperature $T_c$ as functions of electron-phonon coupling and on-site repulsion. Comparisons with the bond SSH model on square lattice and Holstein model are also discussed. Finally, Sec.~\ref{sec:sec6} summarizes our main conclusions.

\section{Model}
\label{sec:sec2}

In this work, we consider the bond-coupled SSH model on a two-dimensional triangular lattice, where the electronic hopping between nearest-neighbor sites is modulated by local lattice vibrations associated with oscillators located on the connecting bonds. The total Hamiltonian consists of three parts $H = H_{\rm e} + H_{\rm ph} + H_{\rm e\text{-}ph}$,
representing the electron, phonon, and electron-phonon interaction terms, respectively.

\begin{equation}
H_{\rm e} = -t \sum_{\langle i,j \rangle, \sigma}(c_{j, \sigma}^{\dagger}c_{i, \sigma}^{\,} + h. c.) +U \sum_{i} n_{i,\uparrow} n_{i,\downarrow} ,
\label{He}
\end{equation}
\begin{equation}
H_{\rm ph}= \omega \sum_{b} \bigg{(}b^{\dagger}_{b} b_{b }^{\,} +\frac{1}{2} \bigg{)} ,
\label{Hph}
\end{equation}
\begin{equation}
H_{\rm e-ph}= g  \sum_{b, \sigma} \bigg{(} c_{j, \sigma}^{\dagger} c_{i, \sigma}^{\,} + h. c.  \bigg{)} \bigg{(}b_{b}^{\dagger} + b_{b}^{\,} \bigg{)}.
\label{Heph}
\end{equation}
where $b=\langle i, j \rangle$ label the nearest neighbor lattice sites (defining the bond $b$), respectively. Here $c_{i,\sigma}$ ($b_i$) are the electron (optical phonon) annihilation operators on site $i$ (on bond $b$), $\sigma \in \{ \uparrow, \downarrow \}$ is the spin index. Here, $t$ is the nearest-neighbor hopping amplitude (used as the energy unit), $U$ is the on-site Coulomb repulsion (nonzero only for bipolarons), $\omega$ is the Einstein phonon frequency, and $g$ denotes the strength of the bond electron-phonon coupling. 

For the Holstein model, the electron-phonon interaction takes the form
\begin{equation}
H_{\mathrm{e\text{-}ph}}^{\mathrm{H}} = g_H \sum_{i, \sigma} 
n_{i, \sigma} \left( b_i^\dagger + b_i \right),
\end{equation}
where \( n_{i,\sigma} = c_{i,\sigma}^\dagger c_{i,\sigma} \) is the electron number operator, and \( b_i^\dagger \) (\( b_i \)) creates (annihilates) a phonon at site \( i \). This on-site coupling describes the electron-phonon interaction between the local lattice displacement and the electron density.

The dimensionless electron-phonon coupling parameter for the bond model is defined as
\begin{equation}
\lambda = \frac{g^2}{d t \omega},
\label{lambda}
\end{equation}
which is a standard convention in the polaron literature. Here, $d$ denotes the dimensionality of the system, with $d = 2$ for the two-dimensional triangular lattice considered in this work.

To enable consistent comparison of physical quantities such as the effective mass \( m^*/m_0 \) and superfluid transition temperature \( T_c \) between the bond and Holstein models, we adopt the definition \( \lambda = \frac{g_H^2}{8 t \omega} \) for the Holstein case. This ensures that results from both models are presented on a comparable coupling scale in the figures.

\section{Method}
\label{sec:sec3}

To investigate polaron and bipolaron formation on a two-dimensional triangular lattice, we employ the diagrammatic Monte Carlo (DiagMC) method within the lattice path-integral framework~\cite{PhysRevB.105.L020501}. DiagMC is an effective and widely used approach for simulating polaronic systems~\cite{PhysRevB.77.125101,PhysRevLett.81.2514,PhysRevB.77.020408,PhysRevB.62.6317}. This method yields numerically exact, sign-problem-free results for a wide range of model parameters. It enables direct access to ground-state properties, including the polaron ground-state energy, effective mass, average phonon number, as well as bipolaron properties such as binding energy and mean-squared radius.


The triangular lattice, characterized by a higher coordination number ($z = 6$) compared to the square lattice ($z = 4$), provides more available hopping pathways for electrons. In the bond SSH model, where the electron-phonon interaction modulates the hopping amplitudes through phonon modes on the bonds, the denser bond network of the triangular lattice amplify the coupling effects. This enhanced connectivity facilitates greater electron mobility and more efficient electron-phonon coupling, thereby influencing polaron and bipolaron properties such as effective mass and bipolaron size. These microscopic features, in turn, play a critical role in determining the emergent collective behavior—particularly the superfluid transition temperature $T_c$ in the dilute limit. Since our study concerns electron-based systems, $T_c$ may also be referred to as the superconducting transition temperature in this context.

For a generic a few body system with Hamiltonian $\hat{H}$, imaginary-time evolution projects an arbitrary state onto the ground state. We measure the standard $n$-particle Green's Function: $G_{ba}(\tau)$. The decay of the propagator $G_{ba}(\tau)$ at long $\tau$ encodes the ground-state energy $E_g$:
\begin{equation}
G_{ba}(\tau) \xrightarrow[\tau \to \infty]{} \langle b | g \rangle \langle g | a \rangle e^{-\tau E_g}.
\end{equation}
where $|a\rangle$ and $|b\rangle$ are states with nonzero overlap with the ground state $|g\rangle$.

By fitting this exponential decay, we extract $E_g$ for both single-polaron and bipolaron sectors. The binding energy is then defined as
\begin{equation}
\Delta_{\text{BP}}(\mathbf{k}) = 2E_{\text{p}}(\mathbf{k}) - E_{\text{BP}}(\mathbf{k}),
\end{equation}
where $E_{\text{p}}(\mathbf{k})$ and $E_{\text{BP}}(\mathbf{k})$ are the single-polaron and bipolaron energies at momentum $\mathbf{k}$.

To evaluate the polaron effective mass $m^*$, we analyze the spatial broadening of the imaginary-time propagator. For large $\tau$ and relative displacement $\mathbf{R}$ between initial and final center-of-mass coordinates~\cite{PhysRevLett.74.2288}, the Green’s function scales as
\begin{equation}
G_{ba}(\tau, \mathbf{R}) \sim \frac{A_{ba} e^{-\tau E_g}}{\tau^{d/2}} \exp\left(-\frac{m^*R^2}{2 \tau}\right).
\end{equation}
This leads to the estimator
\begin{equation}
\overline{\mathbf{R}^2}(\tau) = \frac{\sum_{ab} W_{ab} G_{ba}(\tau, \mathbf{R}) \mathbf{R}^2 }{\sum_{ab} W_{ab} G_{ba}(\tau, \mathbf{R})} \xrightarrow[\tau \to \infty]{} \frac{d}{m^*} \tau.
\end{equation}
The effective mass is then extracted from the slope of $\overline{\mathbf{R}^2}(\tau)$ in the large-$\tau$ limit.

The mean squared radius of the bipolaron provides a measure of its spatial extent. It is obtained by sampling the interparticle distance $\mathbf{r}$ at the midpoint of long bipolaron trajectories:
\begin{equation}
R^2_{\text{BP}} = \sum_{\mathbf{r}} \left(\frac{r}{2}\right)^2 P(\mathbf{r}),
\end{equation}
where $P(\mathbf{r})$ is the probability distribution for finding the two electrons at separation $\mathbf{r}$ in the ground state.

Compact and dilute bipolarons can be treated as composite bosons undergoing a Berezinskii-Kosterlitz-Thouless (BKT) transition in 2D. The corresponding superfluid transition temperature can be estimated using \cite{PhysRevB.104.L201109, PhysRevLett.130.236001, PhysRevLett.87.270402, PhysRevLett.100.140405}:
\begin{equation}
T_c \approx
\begin{cases}
\displaystyle\frac{C}{m_{\text{BP}}^* R_{\text{BP}}^2}, & \text{if } R_{\text{BP}}^2 \gtrsim 1 \\
\displaystyle\frac{C}{m_{\text{BP}}^*}, & \text{otherwise}
\end{cases}
\end{equation}
where $m_{\text{BP}}^*$ and $R^2_{\text{BP}}$ denote the bipolaron effective mass and mean-squared radius, respectively, and $C \approx 0.5$ is a numerical constant appropriate for dilute bosonic systems in two dimension. This expression provides an upper bound for $T_c$ that remains accurate even in the presence of interactions between bosons, such as onsite or long-range Coulomb repulsion.

All calculations are carried out on a two-dimensional triangular lattice with $L = 140$ sites along each principal lattice direction and open boundary conditions. The oblique geometry of the lattice is fully preserved, and finite-size effects have been checked and found to be negligible.

\begin{figure}[t]
\includegraphics[width=0.45\textwidth]{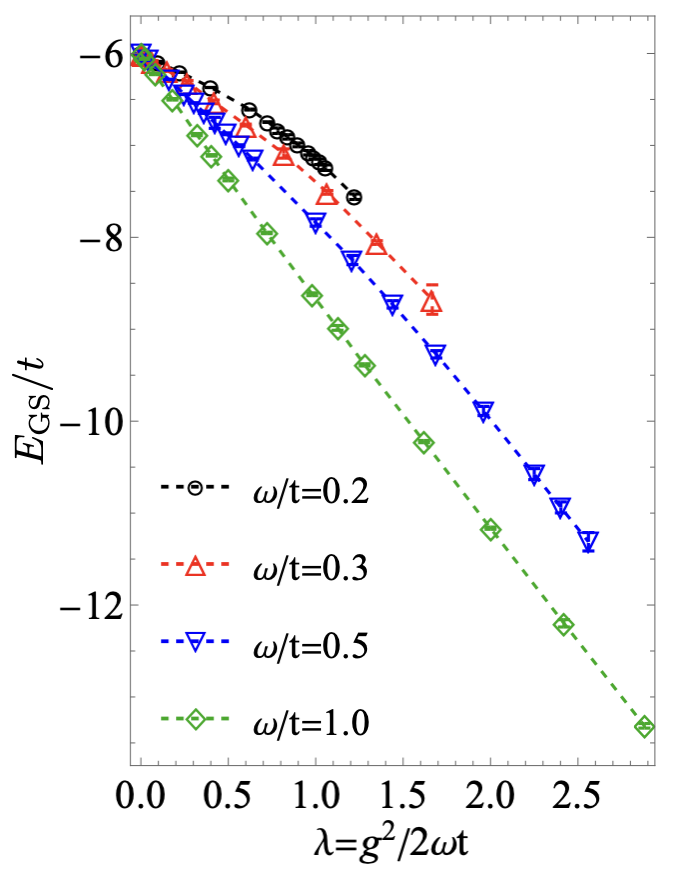}
\caption{Ground-state energy $E_{\mathrm{GS}}/t$ of the bond polaron as a function of the dimensionless coupling $\lambda = g^2 /2 \omega t$, for phonon frequencies $\omega/t = 1.0$ (black circles), $0.5$ (red up-triangles), $0.3$ (blue down-triangles), and $0.2$ (green diamonds). The results illustrate a smooth decrease of $E_{\mathrm{GS}}/t$ with increasing $\lambda$, with more pronounced nonlinearity and slope change emerging at lower $\omega/t$.
} 
\label{FIG1}
\end{figure}

\begin{figure}[b]
\includegraphics[ width=0.45\textwidth]{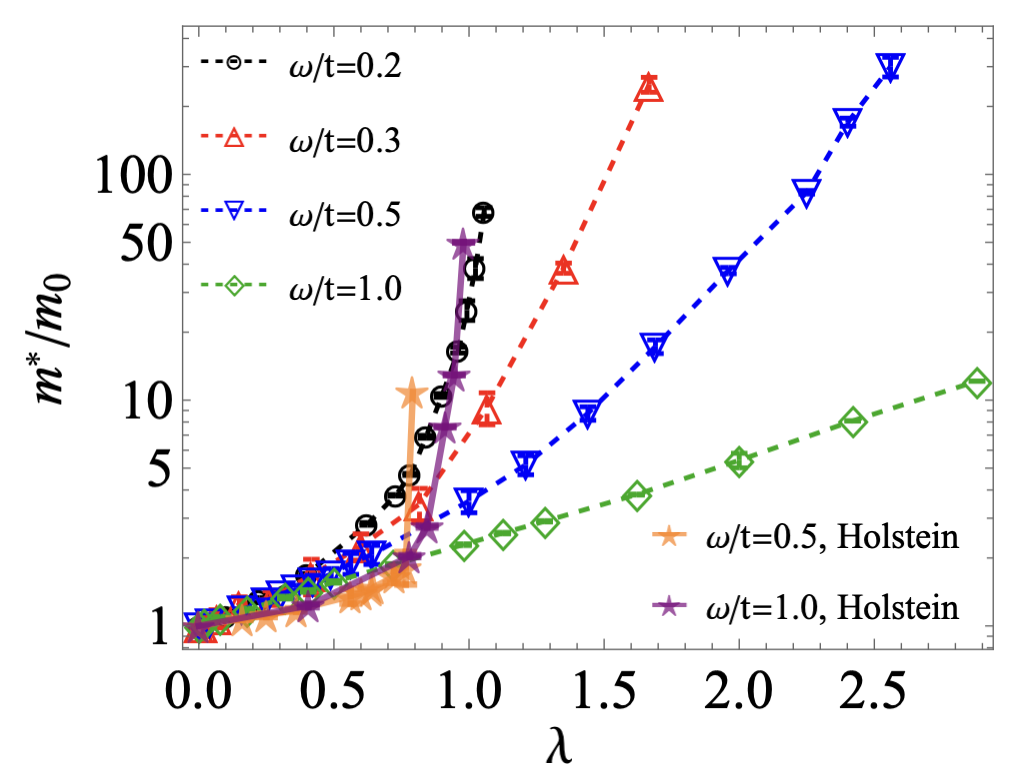}
\caption{Effective mass $m^*/m_0$ of the bond polaron as a function of dimensionless electron-phonon coupling $\lambda = g^2 / 2\omega t$ for various phonon frequencies $\omega/t = 1.0$, $0.5$, $0.3$, and $0.2$. Here, $m_0 = 1/3t$ denotes the mass of a free electron on the triangular lattice. For comparison, Holstein polaron results are shown using $\lambda = g_H^2 / 8\omega t$ for $\omega/t = 1.0$ and $0.5$ (solid stars). As $\omega/t$ decreases, the crossover from a weakly dressed to a strongly dressed polaron shifts to smaller $\lambda$, with an increasingly rapid growth in $m^*/m_0$. In contrast, Holstein polarons exhibit a sharper onset of exponential mass enhancement, reflecting stronger localization due to on-site coupling.}
\label{FIG2}
\end{figure}

\begin{figure}[t]
\includegraphics[width=0.45\textwidth]{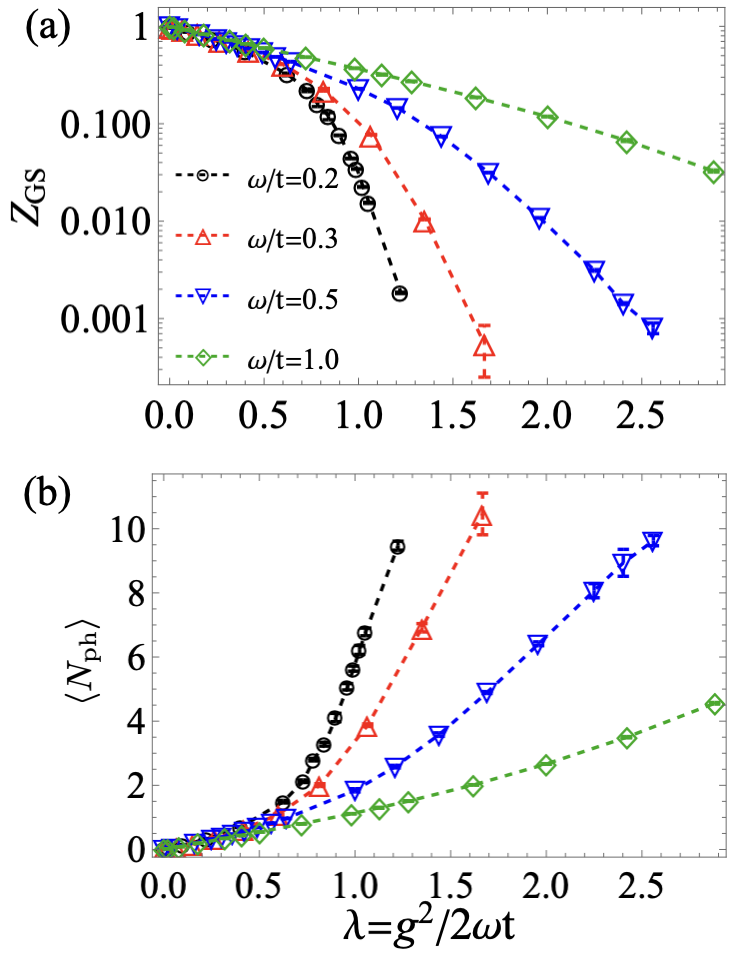}
\caption{
(a) Quasiparticle residue $Z$ of the bond polaron as a function of dimensionless electron-phonon coupling $\lambda$, for phonon frequencies $\omega/t = 1.0$, $0.5$, $0.3$, and $0.2$. $Z$ decreases monotonically with increasing $\lambda$, with a more rapid suppression observed at lower phonon frequencies, reflecting stronger electron-phonon renormalization in the adiabatic regime. 
(b) Average phonon number $\langle N_{\mathrm{ph}} \rangle$ of the bond polaron as a function of $\lambda$, for the same set of phonon frequencies. The growth of $\langle N_{\mathrm{ph}} \rangle$ accelerates at lower $\omega/t$, consistent with enhanced lattice distortion and heavier phonon dressing in the deep adiabatic limit.
}
\label{FIG3}
\end{figure}

\section{The property of a single bond polaron}
\label{sec:sec4}

In this section, we investigate the properties of a single bond polaron on a two-dimensional triangular lattice, including its ground-state energy, effective mass, quasiparticle residue ($Z$ factor), and the average phonon number. In particular, we analyze how lattice geometry influences these properties, with a focus on the effective mass, which plays a critical role in determining the superconducting transition temperature in the bond bipolaron model. Achieving a high transition temperature requires the formation of light bipolarons at reasonable electron-phonon coupling strengths. Since the effective mass of the polaron sets a fundamental limit on the bipolaron mass, a lighter polaron enhances the prospects for realizing light bipolarons and, consequently, higher transition temperatures.

Figures~\ref{FIG1} to \ref{FIG3} present the main results of our study on the bond polaron in the deep adiabatic regime $\omega/t \le 1.0$. We examine the ground-state energy $E_{\mathrm{GS}}/t$ [Fig.\ref{FIG1}], the effective mass $m^*/m_0$ [Fig.\ref{FIG2}], the quasiparticle residue $Z$ factor [Fig.\ref{FIG3}(a)], and the average phonon number $\langle N_{\mathrm{ph}} \rangle$ [Fig.\ref{FIG3}(b)] as functions of the dimensionless electron-phonon coupling $\lambda=g^2/2\omega t$. The calculations are performed for phonon frequencies $\omega/t = 1.0$ (black circles), $0.5$ (red up-triangles), $0.3$ (blue down-triangles), and $0.2$ (green diamonds), allowing us to explore the evolution of polaron properties across different adiabaticity regimes.

Figure~\ref{FIG1} shows that the ground-state energy $E_{\mathrm{GS}}/t$ decreases smoothly as the dimensionless coupling $\lambda$ increases. For higher phonon frequencies ($\omega/t = 1.0$), $E_{\mathrm{GS}}/t$ exhibits an approximately linear dependence on $\lambda$ over a broad range. As the phonon frequency decreases, the $E_{\mathrm{GS}}/t$ curves become increasingly nonlinear, characterized by steeper descent and changes in slope, signaling stronger phonon dressing and more pronounced lattice deformation in response to the electron. In particular, for $\omega/t = 0.2$, a distinct change in slope appears even at relatively small $\lambda$, indicating a crossover to a self-trapped polaron regime. In this deep adiabatic limit, the electron becomes localized due to its strong coupling to a significantly distorted local lattice configuration, resulting in a substantial reduction of the total energy.

\begin{figure}[t]
\includegraphics[width=0.45\textwidth]{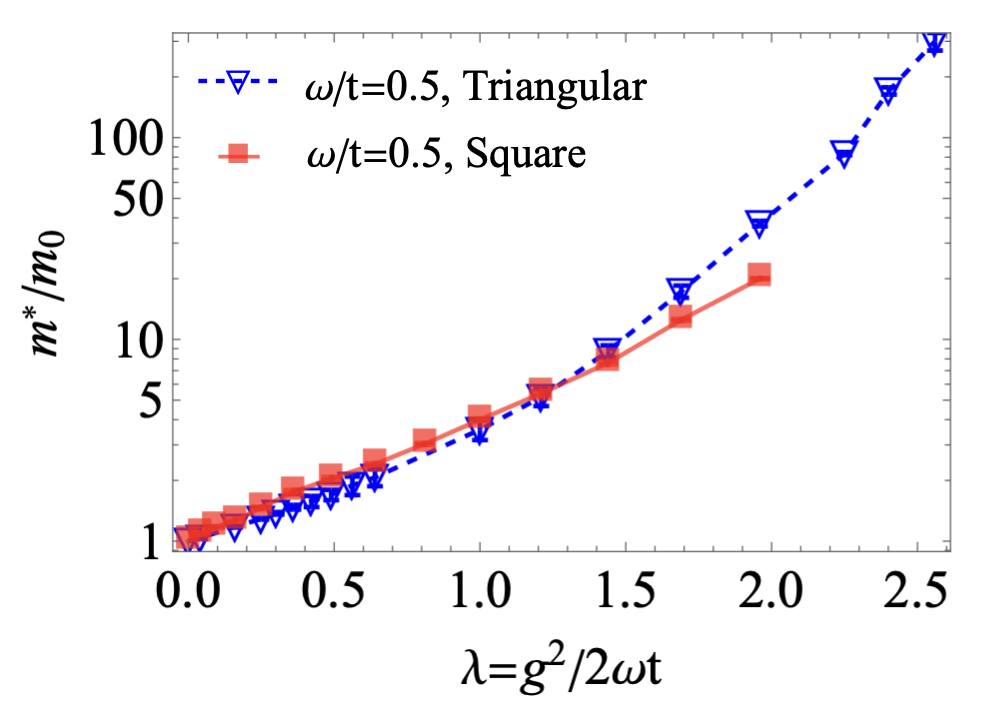}
\caption{Comparison of the effective mass $m^*/m_0$ of bond polarons on square ($m_0=1/2t$) and triangular lattices ($m_0=1/3t$) at phonon frequency $\omega/t = 0.5$. The square lattice data is taken from Ref.~\cite{PhysRevB.104.035143}. For $\lambda \le 1.2$, the triangular lattice exhibits a lower effective mass, indicating enhanced polaron mobility, while for $\lambda \ge 1.2$, the effective mass on the triangular lattice surpasses that of the square lattice. The lighter effective mass on the triangular lattice in the weak to intermediate coupling regime ($\lambda \le 1.0$) supports the emergence of higher bipolaronic transition temperatures in this regime.}
\label{FIG4}
\end{figure}

\begin{figure*}[t]
\includegraphics[ width=0.7\textwidth]{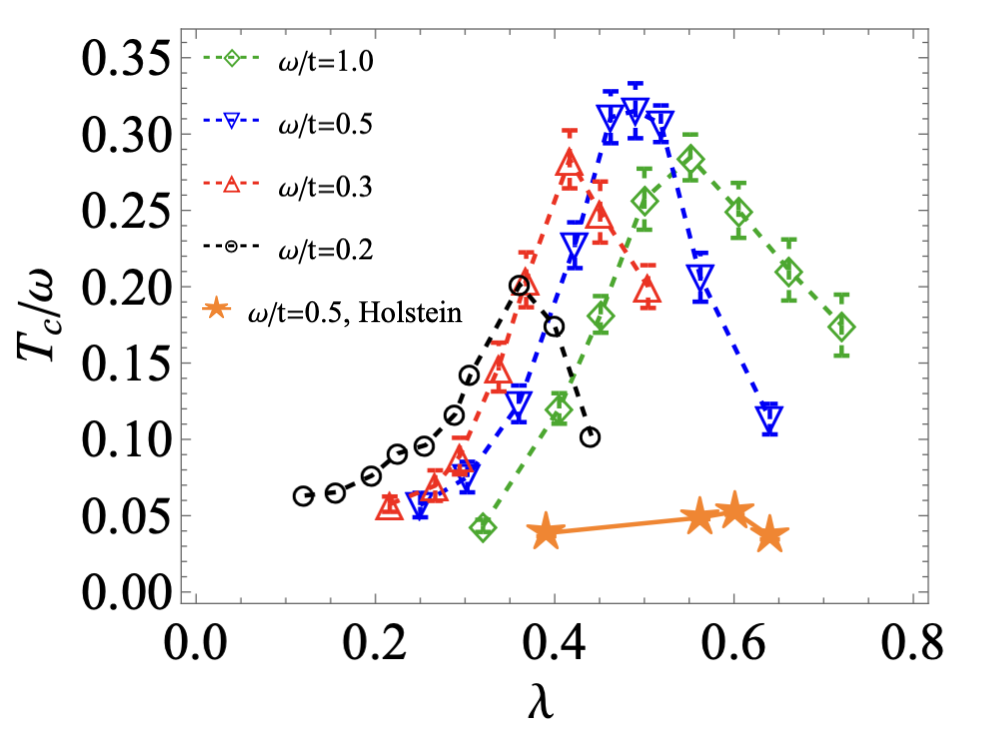}
\caption{Superfluid transition temperature $T_c/\omega$ as a function of coupling strength $\lambda$ for phonon frequencies $\omega/t = 1.0$ (green diamonds), $0.5$ (blue downward triangles), $0.3$ (red upward triangles), and $0.2$ (black circles), at fixed on-site repulsion $U/t = 6.0$. For all $\omega/t$, $T_c$ exhibits a broad maximum that shifts to higher $\lambda$ as $\omega/t$ increases. The maximum $T_c$ values on the triangular lattice exceed those previously reported for the square lattice~\cite{PhysRevX.13.011010}, even in the deep adiabatic regime. For comparison, the $T_c/\omega$ of Holstein bipolarons on the triangular lattice is also shown, using $\lambda = g_H^2 / (8 \omega t)$; for $\omega/t = 0.5$, the Holstein $T_c/\omega$ remains below 0.05.
}
\label{FIG5}
\end{figure*}

The evolution of the effective mass $m^*/m_0$ as a function of the electron-phonon coupling strength $\lambda$ is shown in Fig.~\ref{FIG2}. Here, $m_0 = 1/3t$ denotes the mass of a free electron on the triangular lattice, where the tight-binding dispersion yields an effective mass $m_0$ near the bottom of the band. For all phonon frequencies, $m^*/m_0$ increases with $\lambda$, reflecting enhanced phonon dressing. To enable a direct comparison between bond SSH and Holstein polarons, we adopt the conventions $\lambda = g^2 /2 \omega t$ and $\lambda = g_H^2 /8 \omega t$, respectively. In the bond SSH model, at $\omega/t = 1.0$, the mass enhancement is approximately linear up to $\lambda \approx 2.5$, indicating weak dressing. As $\omega/t$ decreases, this crossover to heavy-polaron behavior shifts to smaller $\lambda$. At $\omega/t = 0.2$, for example, the effective mass increases rapidly beyond $\lambda \approx 0.7$, approaching an exponential growth regime, signaling strong localization in the deep adiabatic limit due to substantial lattice deformation. The Holstein polaron results, shown for $\omega/t = 1.0$ and $0.5$, display a more abrupt onset of exponential mass enhancement, highlighting the stronger localization induced by on-site coupling. Our Holstein polaron results agree well with previous studies~\cite{PhysRevB.73.054303}, validating the accuracy of the simulation method.

Figure~\ref{FIG3}(a) shows the quasiparticle residue $Z$ as a function of the electron-phonon coupling strength $\lambda$ for various phonon frequencies $\omega/t = 1.0$, $0.5$, $0.3$, and $0.2$. For all cases, $Z$ decreases gradually at small $\lambda$, followed by a smooth yet noticeable change in slope over an intermediate range of $\lambda$. This crossover region reflects the gradual loss of quasiparticle residue as the electron becomes increasingly dressed by lattice distortions. The decline in $Z$ is more rapid for lower $\omega/t$, consistent with the heavier polaron effective mass observed in Fig.~\ref{FIG2}, indicating stronger electron-phonon renormalization in the deep adiabatic regime. Figure~\ref{FIG3}(b) presents the average phonon number $\langle N_{\mathrm{ph}} \rangle$ associated with the bond polaron. As $\lambda$ increases, $\langle N_{\mathrm{ph}} \rangle$ grows approximately linearly at first, but then exhibits an accelerated increase in a similar crossover region. This behavior reflects the growing involvement of phonons in polaron formation, as the lattice deformation becomes more significant. The trend is again more pronounced at lower phonon frequencies, confirming that the polaron becomes more heavily dressed and localized in the deep adiabatic limit.

Figure~\ref{FIG4} compares the effective mass $m^*/m_0$ of the bond polaron on square and triangular lattices at phonon frequency $\omega/t = 0.5$. The data for the square lattice is taken from Ref.~\cite{PhysRevB.104.035143}. Here, $m_0$ denotes the mass of a free electron on each lattice. For the triangular lattice, $m_0 = 1/3t$, while for the square lattice, $m_0= 1/2t$. These values reflect the different curvatures of the tight-binding energy bands near the bottom, due to the distinct lattice geometries. All effective mass values $m^*/m_0$ are presented in units normalized to their respective $m_0$, allowing meaningful comparisons between different lattices. 
At weak to intermediate coupling strengths ($\lambda \leq 1.2$), the effective mass on the triangular lattice is lower than that on the square lattice. This lighter mass arises from the triangular lattice’s higher coordination number ($z = 6$) and the increased number of bonds per site, which offer more available hopping pathways. In the bond SSH model, where the electron-phonon interaction modulates hopping amplitudes via bond-centered phonons, this enhanced connectivity amplifies the impact of electron-phonon coupling and facilitates polaron delocalization. As a result, electrons are less localized and acquire smaller mass renormalization in the triangular geometry under weak to moderate coupling. However, in the strong coupling regime ($\lambda \gtrsim 1.2$), this trend reverses: the effective mass on the triangular lattice becomes larger than that on the square lattice. This crossover reflects the fact that in the deep adiabatic limit, the same bond-rich environment that aids mobility at small $\lambda$ now enhances the extent of lattice deformation around the polaron as $\lambda$ increasing. The increased number of phonon-active bonds per site leads to a larger phonon cloud that must move with the electron, producing a heavier dressed quasiparticle. Consequently, the triangular lattice, while favorable for light polarons at small $\lambda$, exhibits stronger localization-induced mass enhancement at large $\lambda$ due to its more extensive lattice coupling. Importantly, within the physically relevant regime $\lambda < 1.0$, the triangular lattice maintains a lighter effective mass, offering favorable conditions for realizing higher bipolaron transition temperatures in this geometry, see Section~\ref{sec:sec5}.

Taken together, these results provide a comprehensive picture of how polaron properties are governed by both the electron-phonon coupling strength and the phonon frequency. In particular, the polaron effective mass, which directly controls its mobility and, consequently, the superfluid or superconducting transition temperature in the bond bipolaron model, is found to be sensitive to lattice geometry. The emergence of light polarons at moderate coupling strengths thus appears essential for achieving high transition temperatures in the bond-SSH bipolaron framework.

\section{Bond bipolaron and high $T_c$ superconductivity}
\label{sec:sec5}

\begin{figure}[b]
\includegraphics[width=0.45\textwidth]{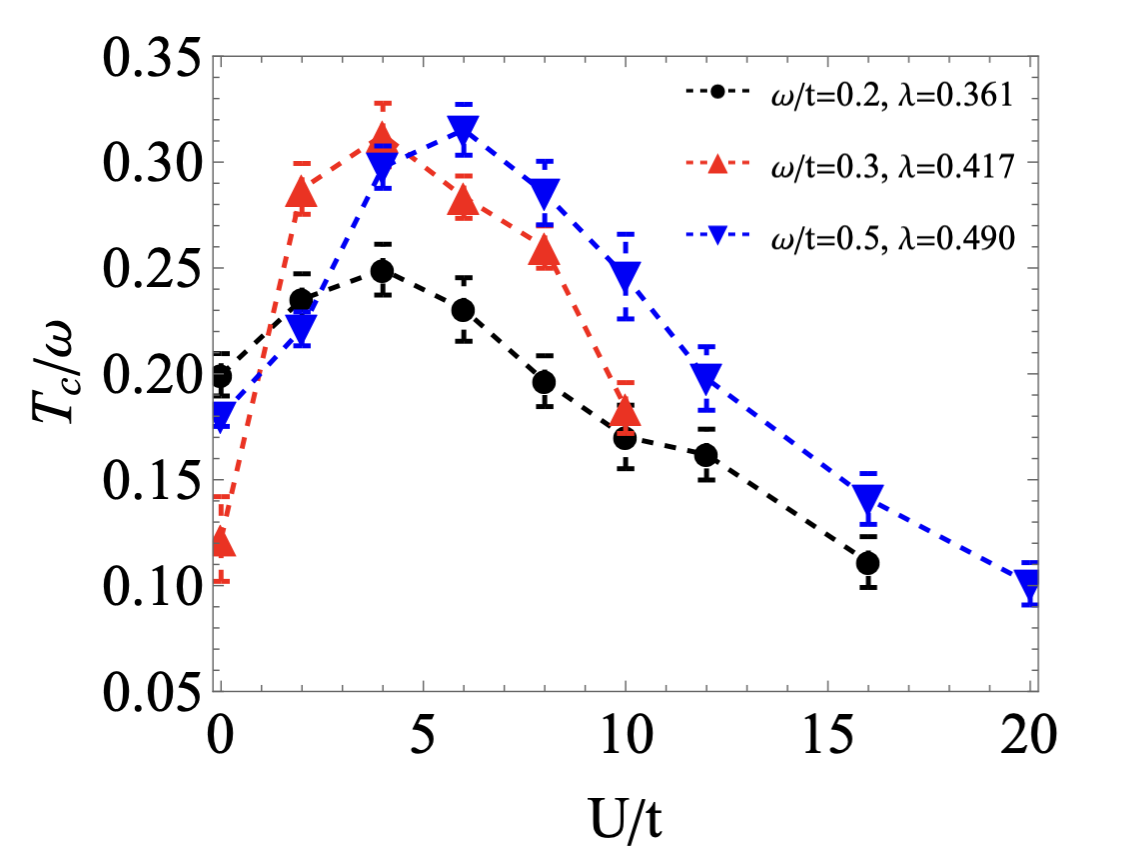}
\caption{Superfluid transition temperature $T_c/\omega$ as a function of on-site interaction strength $U/t$ for $\omega/t = 0.5$ with $\lambda = 0.49$, $\omega/t = 0.3$ with $\lambda = 0.417$, and $\omega/t = 0.2$ with $\lambda = 0.361$. The maximum $T_c/\omega$ occurs at $U/t \approx 4.0$–$6.0$, indicating that moderate on-site repulsion enhances superfluidity by optimizing bipolaron properties.}
\label{FIG6}
\end{figure}

\begin{figure}[h]
\includegraphics[width=0.45\textwidth]{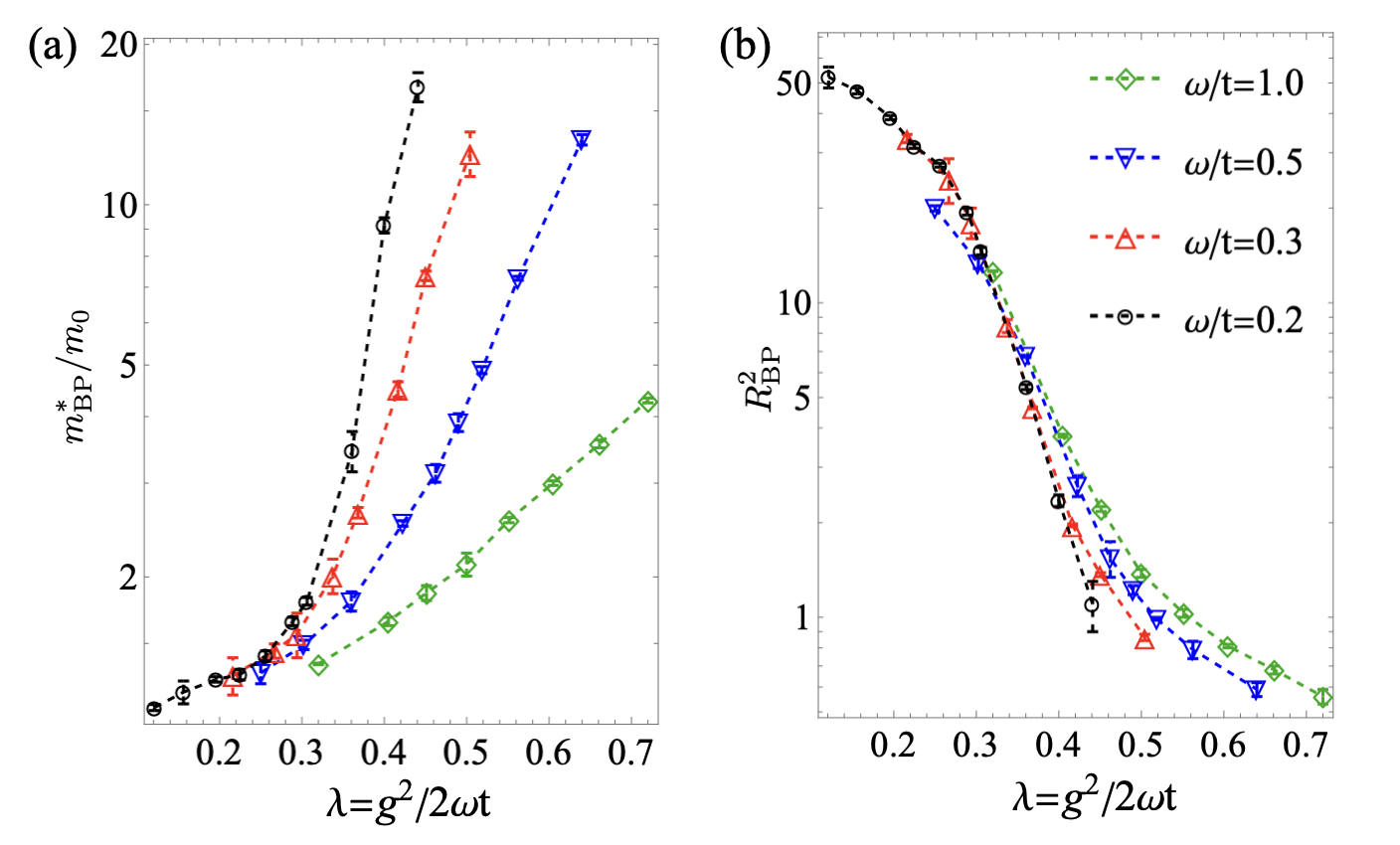}
\caption{Bipolaron properties as a function of electron-phonon coupling $\lambda$, for phonon frequencies $\omega/t = 1.0$, $0.5$, $0.3$, and $0.2$. (a) Bipolaron effective mass $m^*_{\text{BP}}/m_0$ in units of $m_0 = 2 m_e =2/3$, where $m_0$ is the bare mass of two electrons in the triangular lattice. (b) Mean-squared bipolaron radius $R_{\text{BP}}^2$ as a function of $\lambda$ for different phonon frequencies.}
\label{FIG7}
\end{figure}

\begin{figure}[b]
\includegraphics[ width=0.45\textwidth]{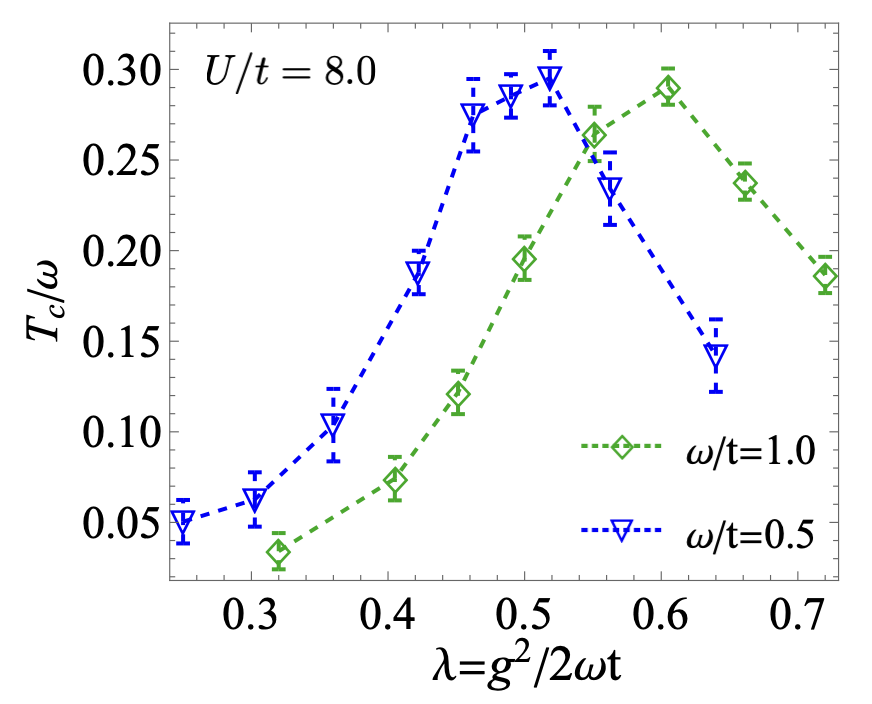}
\caption{Superfluid transition temperature $T_c/\omega$ as a function of electron-phonon coupling strength $\lambda$ for phonon frequencies $\omega/t = 0.5$ and $1.0$ at fixed on-site interaction $U/t = 8.0$. }
\label{FIG8}
\end{figure}

\begin{figure*}[t]
\includegraphics[ width=0.9\textwidth]{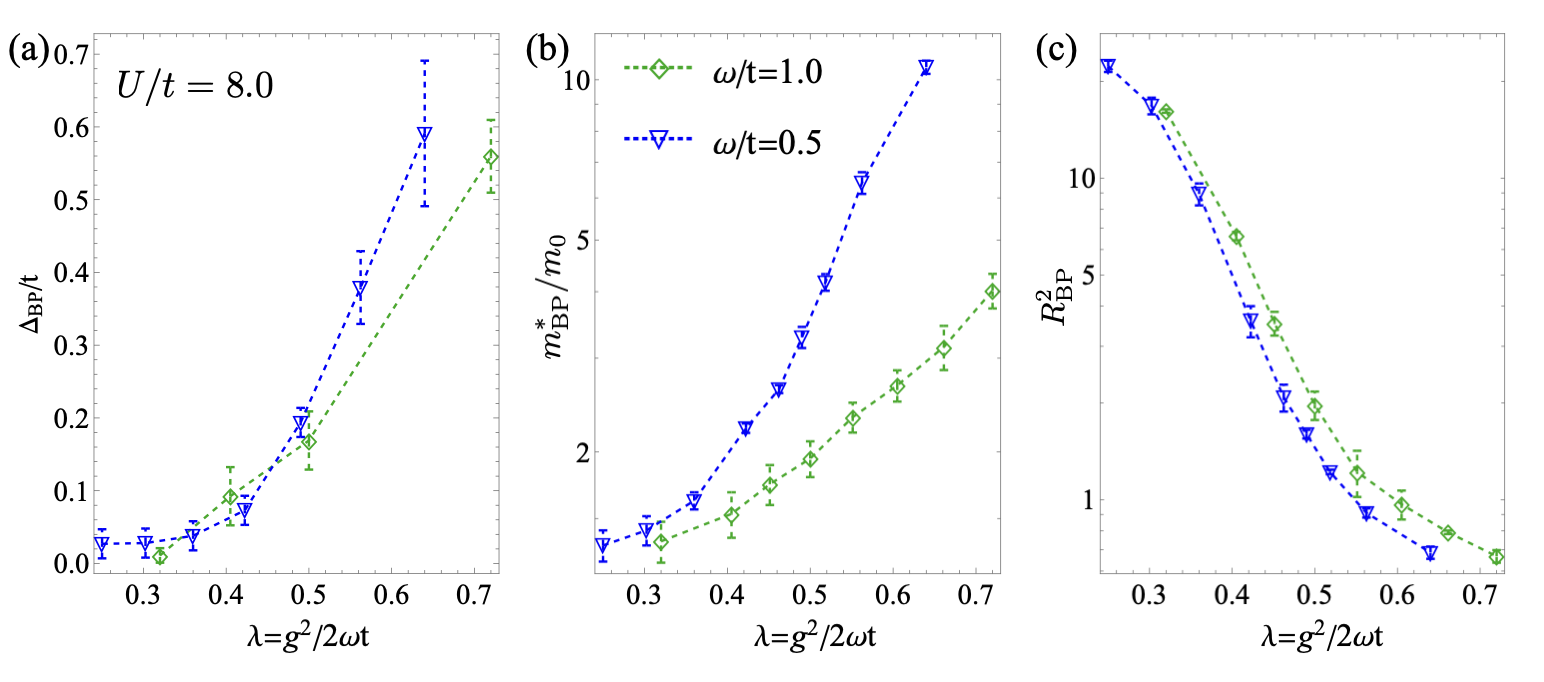}
\caption{(a) Bipolaron binding energy, (b) effective mass $m^*_{\text{BP}}/m_0$, and (c) mean-squared radius $R_{\text{BP}}^2$ as functions of electron-phonon coupling $\lambda$ for $\omega/t = 0.5$ and $1.0$ at $U/t = 8.0$. }
\label{FIG9}
\end{figure*}

Building on the analysis of single-polaron properties, we now turn to the study of bond bipolarons on the triangular lattice. In this section, we investigate their binding energy, effective mass, and mean-squared radius, as well as the resulting superconducting transition temperature $T_c$ in the dilute limit. Our analysis highlights how these bipolaron properties evolve with electron-phonon coupling strength, phonon frequency, and on-site Hubbard interaction.

Figure~\ref{FIG5} presents the superfluid transition temperature $T_c/\omega$ in two dimensional triangular lattice as a function of coupling strength $\lambda$, for phonon frequencies $\omega/t = 1.0$ (green diamonds), $0.5$ (blue downward triangles), $0.3$ (red upward triangles), and $0.2$ (black circles), at fixed on-site repulsion $U/t = 6.0$. The choice of $U/t = 6.0$ is motivated by the results in Fig.~\ref{FIG6}, where this value yields the maximum $T_c/\omega \sim 0.32$ for $\omega/t = 0.5$. For all phonon frequencies considered, $T_c/\omega$ exhibits a broad maximum as a function of $\lambda$. The peak $T_c/\omega$ occurs around $\lambda \approx 0.3$ for $\omega/t = 1.0$, $0.5$, and $0.3$, indicating that moderate coupling strengths optimize superconducting temperature in this regime. Notably, the maximum $T_c/\omega$ values obtained on the triangular lattice are higher than those previously reported for the square lattice (around 0.2)~\cite{PhysRevX.13.011010}, even in the adiabatic regime. In particular, for $\omega/t = 0.2$, the triangular lattice still supports relatively high value of $T_c/\omega \sim 0.25$, comparable to or exceeding the highest values reported for the square lattice~\cite{}. This directly highlights the geometric advantage of the triangular lattice in enhancing bipolaron mobility and pairing in the bond SSH electron-phonon coupling. 

To facilitate direct comparison with Holstein bipolarons, we also plot $T_c/\omega$ as a function of the dimensionless coupling $\lambda = g_H^2 / (8 \omega t)$, where $g_H$ is the Holstein coupling strength, on the same figure. For $\omega/t = 0.5$, the superfluid transition temperature of Holstein bipolarons on the triangular lattice remains below 0.05, significantly lower than that of bond bipolarons. This contrast highlights the physical advantage of bond bipolarons: in the bond SSH model, phonons couple to the electron hopping amplitudes along bonds, making the interaction highly sensitive to lattice connectivity. The triangular lattice, with its higher coordination number ($z = 6$) and denser bond network, enhances carrier mobility and facilitates the formation of light, compact bipolarons. As a result, bond bipolarons exhibit markedly higher $T_c/\omega$ values than their Holstein counterparts, which are limited by on-site electron-phonon coupling and less responsive to lattice geometry.

Figure~\ref{FIG6} shows the superfluid transition temperature $T_c/\omega$ as a function of on-site repulsion $U/t$ for three representative parameter sets: $\omega/t = 0.5$ with $\lambda = 0.49$, $\omega/t = 0.3$ with $\lambda = 0.417$, and $\omega/t = 0.2$ with $\lambda = 0.361$. The results show that for $\omega/t = 0.2$, the maximum $T_c/\omega$ reaches approximately $0.25$ at $U/t=4.0$, while for $\omega/t = 0.3$, it reaches approximately $0.3$ at $U/t=4.0$. For $\omega/t = 0.5$, the maximum $T_c/\omega$ is observed at $U/t = 6.0$. In all cases, the optimal $U/t$ lies in the range of $U/t = 4.0$ to $6.0$. These findings demonstrate that a moderate on-site repulsion is beneficial for optimizing $T_c/\omega$, as it enables a balance between bipolaron size and effective mass.

The role of on-site repulsion $U$ is twofold. In the absence of $U$, the two electrons forming a bipolaron tend to localize on the same lattice site, resulting in a tightly bound but very heavy composite particle. This strong localization leads to a large effective mass and reduced mobility, thereby suppressing the superfluid transition. Introducing a moderate $U$ prevents complete on-site localization, encouraging the electrons to partially separate in space. Consequently, the bipolaron becomes more extended and acquires a lower effective mass. This leads to an increase in the transition temperature. Hence, an optimal range of $U$ balances bipolaron effective mass and its size, maximizing $T_c$.

Figure~\ref{FIG7} presents the bipolaron properties of the bond-SSH model as a function of electron-phonon coupling strength $\lambda$, for phonon frequencies $\omega/t = 1.0$, $0.5$, $0.3$, and $0.2$. Panel (a) shows the bipolaron effective mass $m^*_{\text{BP}}/m_0$, where $m_0 = 2* t/3$ is the mass of two free electrons on triangular lattice. For higher phonon frequencies (e.g., $\omega/t = 1.0$), the effective mass increases gradually with $\lambda$ and remains relatively light over a broad range. In contrast, for lower $\omega/t$, especially $\omega/t = 0.2$, the effective mass increases sharply beyond $\lambda \approx 0.25$, indicating strong lattice dressing and a crossover to heavy, less mobile bipolarons in the deep adiabatic regime. Panel (b) shows the corresponding mean-squared bipolaron radius $R_{\text{BP}}^2$. As $\lambda$ increases, $R_{\text{BP}}^2$ decreases across all phonon frequencies. This contraction is more pronounced at lower $\omega/t$, indicating that bipolarons become more tightly bound in the deep adiabatic limit. The competition between enhanced pairing strength (manifested as reduced bipolaron size) and reduced mobility (due to increased effective mass) plays a central role in determining the maximum achievable superfluid transition temperature $T_c$ in bipolaronic systems.

Figure~\ref{FIG8} shows the superfluid transition temperature $T_c$ as a function of electron-phonon coupling $\lambda$ for phonon frequencies $\omega/t = 0.5$ and $1.0$ at fixed on-site repulsion $U/t = 8.0$. Figure~\ref{FIG9} presents the associated bipolaron properties: (a) the bipolaron binding energy, (b) the effective mass $m^*_{\text{BP}}/m_0$, and (c) the mean-squared bipolaron radius $R^2_{\text{BP}}$.

The overall behavior at $U/t = 8.0$ closely resembles that at $U/t = 6.0$, with the system still exhibiting relatively high superfluid transition temperatures. For instance, at $\omega/t = 0.5$, $T_c/\omega$ reaches $\sim 0.3$ near the optimal coupling $\lambda$, demonstrating that moderately strong on-site repulsion does not suppress superfluidity. This elevated $T_c/\omega$ is associated with the formation of lightweight yet tightly bound bipolarons, as reflected in the moderate effective mass and compact size shown in Fig.~\ref{FIG9}(b) and (c). These findings reinforce the conclusion that bond bipolarons on the triangular lattice remain lightweight and spatially compact across a broad range of coupling and interaction strengths, providing favorable conditions for bipolaronic superconductivity.

\section{Conclusion}
\label{sec:sec6}

A comprehensive numerical analysis of bond polarons and bipolarons on a two-dimensional triangular lattice was studied using DiagMC method. The triangular lattice, with its larger coordination number, supports the formation of light polarons and mobile bipolarons across a wide parameter regime. High superfluid transition temperatures $T_c$ emerge when bipolarons are compact and lightweight. A moderate on-site repulsion $U/t$ enhances $T_c$ by optimizing this balance. For example, $T_c/\omega \sim 0.3$ is reached at $\omega/t = 0.5$ and $U/t = 6.0$, with similarly high values $T_c/\omega \sim 0.25$ observed even for $\omega/t = 0.2$ and $U/t=4.0$. Compared to square lattices, the triangular geometry yields consistently higher $T_c$, highlighting the role of lattice geometry in stabilizing lightweight bipolarons. These results demonstrate that bond-centered electron-phonon coupling on triangular lattice can enhance bipolaronic superfluidity, offering a potential route toward high-$T_c$ in engineered quantum materials.

\begin{acknowledgments}
CZ acknowledges support from the National Natural Science Foundation of China (NSFC) under Grants No 12204173 and 12275002, and the University Annual Scientific Research Plan of Anhui Province under Grant No. 2022AH010013. 
\end{acknowledgments}

\bibliography{Dispersive}

\end{document}